\documentclass[aps,preprint]{revtex4}%
\usepackage{amsfonts}
\usepackage{amsmath}
\usepackage{amssymb}
\usepackage{graphicx}
\usepackage{float}%
\providecommand{\U}[1]{\protect\rule{.1in}{.1in}}

\ifx\pdfoutput\relax\let\pdfoutput=\undefined\fi
\newcount\msipdfoutput
\ifx\pdfoutput\undefined\else
\ifcase\pdfoutput\else
\ifx\paperwidth\undefined\else
\ifdim\paperheight=0pt\relax\else\pdfpageheight\paperheight\fi
\ifdim\paperwidth=0pt\relax\else\pdfpagewidth\paperwidth\fi
\fi\fi\fi
\begin{document}
\title{Geometry of the adiabatic theorem}
\author{Augusto C\'{e}sar Lobo}
\affiliation{Departamento de F\'{i}sica - Instituto de Ci\^{e}ncias Exatas e Biol\'{o}gicas - Universidade Federal de Ouro Preto, CEP 35400-000, Ouro Preto - Minas Gerais, Brazil}
\author{Rafael Antunes Ribeiro}
\affiliation{Centro Federal de Educa\c{c}\~{a}o Tecnol\'{o}gica de Minas Gerais Campus VIII - CEP 37022-560, Varginha, Minas Gerais, Brazil }
\author{Clyffe de Assis Ribeiro}
\affiliation{Departamento de F\'{i}sica, Instituto de Ci\^{e}ncias Exatas, Universidade Federal de Minas Gerais, CP 702, CEP 30161-970, Belo Horizonte, Minas Gerais, Brazil}
\author{Pedro Ruas Dieguez}
\affiliation{Departamento de F\'{i}sica, Instituto de Ci\^{e}ncias Exatas, Universidade Federal de Minas Gerais, CP 702, CEP 30161-970, Belo Horizonte, Minas Gerais, Brazil}
\begin{abstract}
We present a simple and pedagogical derivation of the quantum adiabatic
theorem for two level systems (a single qubit) based on geometrical structures
of quantum mechanics developed by Anandan and Aharonov, among others. We have
chosen to use only the minimum geometric structure needed for an understanding
of the adiabatic theorem for this case.

\end{abstract}
\date{\today}
\maketitle
\tableofcontents

\section{Introduction}

The quantum mechanical adiabatic theorem is one of the most important and
fruitful tools of quantum physics. It was first stated in 1928 by Born and
Fock \cite{BornandFock1928} and set in a more rigorous mathematical foundation
in 1950 by Kato \cite{kato1950adiabatic}. In the mid-eighties, it was linked
in a new and fundamental way with a more geometrical view of quantum mechanics
by the so called Berry's phase or geometric phase
\cite{berry1984,aharonov1987}.

More recently, it has received a renewal of interest because of the advent of
quantum information theory and in particular, the quantum adiabatic
computation model \cite{farhi2000quantum,aharonov2004adiabatic}. In fact, some
authors have questioned the theorem and have tried to look for a more precise
statement and conditions for its consistency and validity
\cite{marzlin2004inconsistency,wu2005validity,tong2005quantitative,ma2006comment,wei2007quantum,amin2009consistency,boixo2010necessary,frasca2011consistency,rigolin2011adiabatic}%
.

Though the theorem can be stated and understood quite easily and it is, in
fact, routinely discussed in undergraduate textbooks
\cite{griffiths2005introduction,messiah1962quantum}, we believe that the
reason \textit{why} it works though, is less known. It can be simply stated as
saying that under a certain special context, if the state of a quantum
mechanical system is an eigenstate of its hamiltonian, it will remain as such.
The special context means that the time-dependent hamiltonian must change
slowly in some sense (adiabaticity) if the spectrum obeys some technical
conditions. Stated in this manner, one can say it is a reasonably intuitive
assertion and can be traced back to its correspondent classical theorem on
adiabatic invariants \cite{kasuga1961adiabatic}. Yet, the proofs available to
students such as in \cite{messiah1962quantum} are technically difficult and
for this reason they may not be very enlightening for the average student.

We shall discuss the reasons for the validity of the theorem for two-state
systems in a setting inspired by the geometrical treatment of quantum
mechanics given by Anandan and Aharonov \cite{Anandan1990}. In the next
Section, we review the geometry of quantum evolution for finite dimensional
systems and in particular, for two-level systems (which in modern quantum
information theory are called qubits) and use this to discuss the quantum
adiabatic theorem in a very clean and pedagogical geometric setting which we
believe improves the physical understanding of this important theorem for
students of Physics and even for specialists. In Section III we conclude with
some closing remarks.

\section{The Geometry of Quantum Evolution}

Let $W^{n+1}$ be a $(n+1)$-dimensional Hilbert space together with its dual
$\overline{W}^{n+1}$ and let $\{|u_{\sigma}\rangle\}$ $(\sigma=0,1,...,n)$
also be an arbitrary basis for $W^{n+1}.$ An hermitean inner product may be
introduced by an \textit{anti-linear} mapping $\dag:W^{n+1}\longrightarrow
\overline{W}^{n+1}$ (where $\dag$ is the familiar ``dagger" operation).
Indeed, the inner product between two arbitrary states $|\psi\rangle$ and
$|\varphi\rangle$ can now be defined as%
\[
(|\psi\rangle,|\varphi\rangle)=|\psi\rangle\left(  ^{\dagger}|\varphi
\rangle\right)  =\langle\psi|\varphi\rangle.
\]
Thus, an arbitrary normalized ket $|\psi\rangle$ expanded in such a basis can
be represented by a complex $(n+1)$-column matrix:
\begin{equation}
|\psi\rangle=\sum\limits_{\sigma}|u_{\sigma}\rangle\psi^{\sigma}\equiv(%
\begin{array}
[c]{cccc}%
\psi_{0} & \psi_{1} & ... & \psi_{n}%
\end{array}
)^{\intercal}, \quad\text{with}\quad\sum\limits_{\sigma}\overline{\psi
}_{\sigma}\psi^{\sigma}=1. \label{column matrix expansion of normalized ket}%
\end{equation}
Writing the complex amplitudes as $\psi^{\sigma}=x^{\sigma}+iy^{\sigma}$ one
can easily see that the set of normalized states can be identified with a
$(2n+1)$-dimensional sphere $S^{2n+1}$ contained in $W^{n+1}$. Since two state
vectors that differ by a complex phase cannot be physically distinguished by
any means, it is convenient to define the true physical space of states as the
above set of normalized states \textit{modulo} the equivalence relation in
$S^{2n+1}$ defined in the following way: We say that $|\psi\rangle$ is
equivalent to $|\varphi\rangle$ if and only if there is a real number $\theta$
such that $|\psi\rangle=e^{i\theta}|\varphi\rangle.$ The \textit{space of
rays} defined above is also known as the $n$-dimensional (complex) projective
space $\mathbb{CP}(n)$. A standard complex coordinate system for
$\mathbb{CP}(n)$ is provided by $n$ complex numbers $\xi^{i}=\psi^{i}%
\diagup\psi^{0}$ ($i=1,...,n$) for those points where $\psi^{0}\neq0$. In the
$n=1$ case we have a \textit{single} qubit described by a single complex
coordinate $\xi$. In this case, $\mathbb{CP}(1)$ is topologically equivalent
to a two-dimensional sphere and the stereographic projection map shows
explicitly
\begin{equation}
\xi=\tan(\theta/2)e^{i\varphi} \label{stereographic map}%
\end{equation}
(see the figure below) that $\mathbb{CP}(1)$ can be seen as the complex plane
``plus" a point at\ ``infinity".
\begin{figure}[H]
\label{projecao}
\par
\begin{center}
\includegraphics[scale=.4]{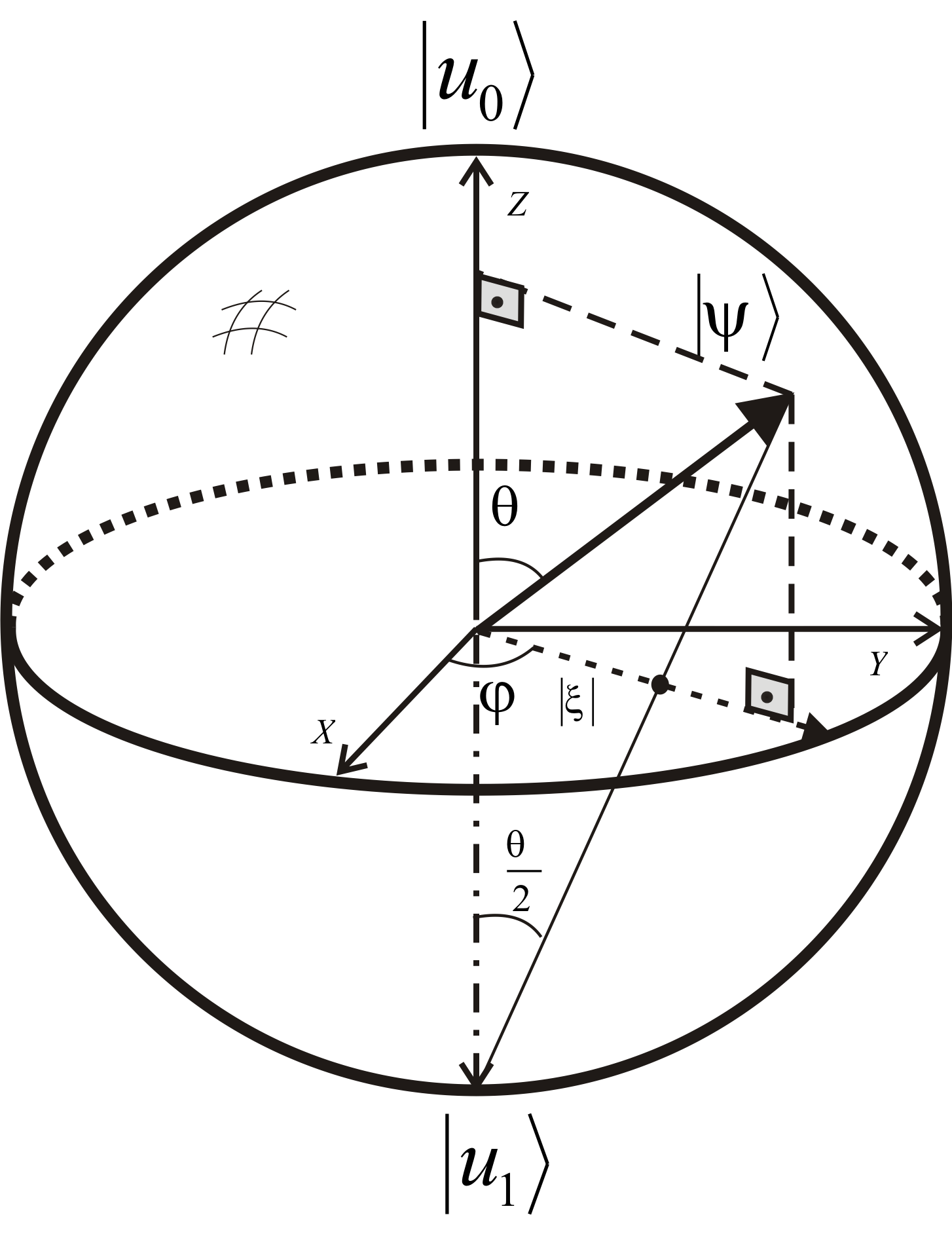}
\end{center}
\caption{Stereographic projection.}%
\end{figure}

This description provides us with the so called Bloch sphere (or Riemann
sphere) with standard coordinates. Thus, any\ physical state can be expressed
as a normalized state represented as a point on the Bloch sphere in the
following standard form
\begin{equation}
|\psi\rangle=|\theta,\varphi\rangle=\cos\left(  \theta/2\right)  |u_{0}%
\rangle+e^{i\varphi}\sin\left(  \theta/2\right)  |u_{1}\rangle,
\label{qubit on Bloch sphere}%
\end{equation}
where one can easily see that antipodal points in the Bloch sphere represent
orthogonal state vectors. A natural metric $\mathbb{CP}(n)$ can be defined in
the following way: Let points $P_{0}$ and $P_{1}$ $\in$ $\mathbb{CP}(n)$ be
projections respectively from two infinitesimally nearby normalized state
vectors $|\psi\rangle$ and $|\psi+d\psi\rangle$. It is then natural to define
the squared distance between $P_{0}$ and $P_{1}$ as the projection of
$|d\psi\rangle$ in the\ \textquotedblleft orthogonal direction" of
$|\psi\rangle$, that is, the projection given by the projection operator
$\hat{\pi}_{|\psi\rangle}^{\perp}=\hat{I}-|\psi\rangle\langle\psi|$ as shown
in FIG. 2. It is then easy to see that (by the
idempotence of $\hat{\pi}_{|\psi\rangle}^{\perp}$)
\begin{equation}
ds^{2}(\mathbb{CP}(n))=\left\vert \langle d\psi|\hat{\pi}_{|\psi\rangle
}^{\perp}|d\psi\rangle\right\vert ^{2}=\langle d\psi|d\psi\rangle-\langle
d\psi|\psi\rangle\langle\psi|d\psi\rangle. \label{metric 2}%
\end{equation}
\begin{figure}[ptb]
\label{geometry of CP(N)}
\par
\begin{center}
\includegraphics[
natheight=3.220300in,
natwidth=2.262900in,
height=2.7203in,
width=2.7629in
]{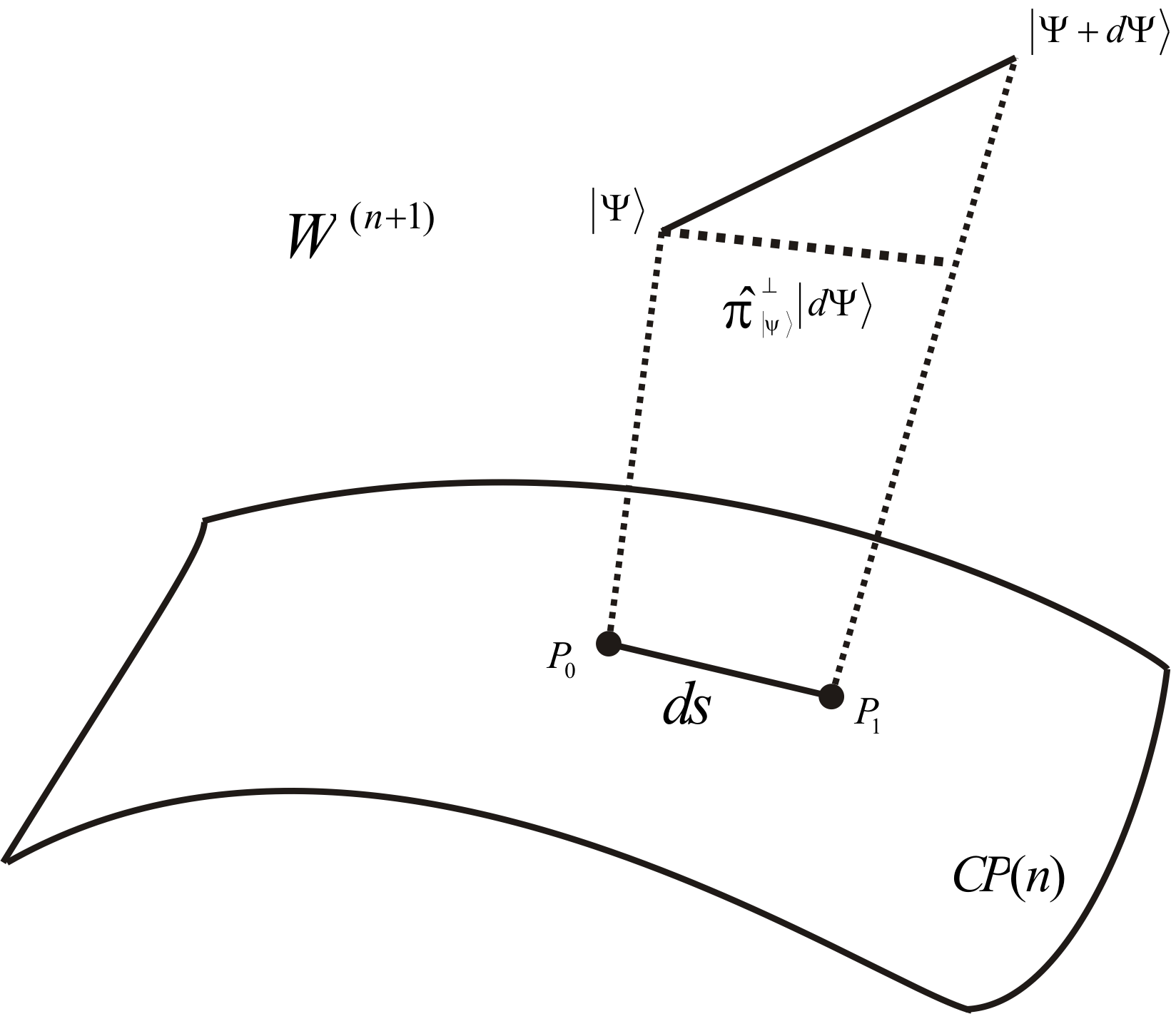}
\end{center}
\caption{Pictorial representation of the geometry in $\mathbb{CP}(n))$}%
\end{figure}

Let $|\psi(t)\rangle$ be the curve of normalized state vectors in $W^{n+1}$
given by the unitary evolution generated by an hamiltonian $\hat{H}$. The
Schr\"{o}dinger equation implies a relation between $|\psi(t)\rangle$ and
$|\psi(t+dt)\rangle$ given by:
\begin{equation}
|d\psi\rangle=|\psi(t+dt)\rangle-|\psi(t)\rangle=-i\hat{H}|\psi(t)\rangle dt,
\label{infinitesimal unitary displacement}%
\end{equation}
for the sake of simplification we have used and henceforth will be assuming
$\hslash=1$ units. The above equation together with (\ref{metric 2}) gives us
the squared distance between two infinitesimally nearby projection of state
vectors connected by the unitary evolution over $\mathbb{CP}(n)$:
\begin{equation}
ds^{2}(\mathbb{CP}(n))=\left[  \langle\psi(t)|\hat{H}^{2}|\psi(t)\rangle
-(\langle\psi(t)|\hat{H}|\psi(t)\rangle)^{2}\right]  dt^{2}=\left(
\delta_{|\psi(t)\rangle}^{2}E\right)  dt^{2}. \label{metric 3}%
\end{equation}

In particular, for a single qubit, one can write the metric over the
$\mathbb{CP}(1)$ sphere by (\ref{qubit on Bloch sphere}) as%
\begin{equation}
ds^{2}=\frac{1}{4}(d\theta^{2}+\sin^{2}\theta d\varphi^{2})
\label{metric on Bloch sphere}%
\end{equation}
where we can immediately identify the Bloch sphere as a $2D$ sphere with
radius $r=1/2$. Note that some authors define the Bloch sphere as a unit
radius sphere. Our choice seems more natural to us because of the above
geometrical construction and also in order to identify the Bloch sphere with
the so called Riemann sphere.

\subsection{The time-energy uncertainty relation}

Equation (\ref{metric 3})\ leads to a very elegant relation between the speed
of the projection over $\mathbb{CP}(n)$ and the instantaneous energy
uncertainty \cite{Anandan1990}.
\begin{equation}
\frac{ds}{dt}=\delta E(t). \label{speed over CP(n)}%
\end{equation}

The above equation contains in itself the seeds of both a beautiful geometric
formulation of the \textit{time-energy uncertainty relation} and as well as
the \textit{adiabatic theorem}. Indeed, for the latter, it is quite clear that
if a state is initially (let us say for $t=0$) an energy eigenstate, then one
has $\delta E(0)=0$ and the instantaneous speed must then also be null. If one
moves the state very slowly around $\mathbb{CP}(n)$, such that $\frac{ds}%
{dt}\approx0$ then one should have in a self consistent way that $\delta
E(t)\approx0$ for all $t$. But this means that the state \textit{must remain}
an eigenstate for all $t$, which is exactly a statement of the adiabatic
theorem and we shall discuss this issue further in the next Subsection.

Before that, let us close this Subsection by observing how equation
(\ref{speed over CP(n)}) also implies a geometric version of the time-energy
uncertainty relation due to Anandan and Aharonov. Let us assume at first that
the system consists only of a \textit{single qubit}. Let $P_{0}$ and $P_{1}$
be two distinct points on Bloch sphere located on an arbitrary path driven by
a time dependent hamiltonian $\hat{H}(t)$ at instants respectively $t_{0}$ and
$t_{1}$. One can define the \textit{time} \textit{average of the energy
uncertainty} $\overline{\delta E}$\ between $t_{0}$ and $t_{1}$ as
\begin{equation}
\overline{\delta E}=\frac{1}{\Delta t}\int_{t_{0}}^{t_{1}}\delta
E(t)dt\qquad\text{with}\qquad\Delta t=t_{1}-t_{0}
\label{time average of energy uncertainty}%
\end{equation}
With this definition and from (\ref{speed over CP(n)}) it is easy to see that%
\begin{equation}
\Delta s=\overline{\delta E}\Delta t
\end{equation}

Now to derive the time-energy uncertainty relation, we may borrow an
elementary argument from quantum mechanics updated with a more modern flavor
from quantum information. In fact, given a particular state (an arbitrary
point in Bloch sphere) one can physically distinguish it from another state if
and only if this second point is \textit{orthogonal} to it (its antipodal
point on the sphere). In fact, the \textit{indistinguishability} of
\textit{non-orthogonal} state vectors is a basic result in quantum information
together with the \textit{no-cloning} theorem, for instance. The
indistinguishability of non-orthogonal states means that one party (let us say
Alice) cannot reliably transmit information to a second party (call him Bob) from a common
agreed upon alphabet of \textit{non-orthogonal} states because these states
cannot be simultaneously eigenstates of commuting observables. For two-level
systems, this means that for the two points $P_{0}$ and $P_{1}$ to be
physically distinguished, their minimum distance must necessarily be the half
length of a great circle of the Bloch sphere, which is clearly $\pi/2$. So we
may write the time-energy uncertainty relation as
\begin{equation}
\overline{\delta E}\Delta t\geqslant\pi/2
\label{geometric time-energy uncertainty relation}%
\end{equation}

Notice that the above inequality differs from the usual text-book relation by
a factor of $\pi$. This is a very different mathematical picture than the one
exhibited by the celebrated Heisenberg uncertainty relation for position and
momentum. But this is expected because in non-relativistic mechanics,
\textit{time} is an external parameter and \textbf{not} an observable as it
happens to be the case for the position operator. In fact, some authors refer
to the time $\Delta t$ in the inequality above as \textit{passage time}
\cite{brody2003elementary}.

\subsection{The Adiabatic Theorem in $\mathbb{CP}(1)$}

By using (\ref{column matrix expansion of normalized ket}), we can write the
Schr\"{o}dinger equation explicitly in coordinates as%
\begin{equation}%
{\displaystyle\sum\limits_{\nu}}
H_{\nu}^{\mu}\psi^{\nu}=i\dot{\psi}^{\mu}
\label{schrodinger equation in coordinates}%
\end{equation}

For a two-level system (a single qubit) one can write%
\begin{equation}
\left\{
\begin{array}
[c]{c}%
H_{0}^{0}\psi^{0}+H_{1}^{0}\psi^{1}=i\dot{\psi}^{0}\\
H_{0}^{1}\psi^{0}+H_{1}^{1}\psi^{1}=i\dot{\psi}^{1}%
\end{array}
\right.  \qquad\text{with\qquad}\hat{H}(t)=\left(
\begin{array}
[c]{cc}%
H_{0}^{0}(t) & H_{1}^{0}(t)\\
H_{0}^{1}(t) & H_{1}^{1}(t)
\end{array}
\right)
\end{equation}

We can project this equation of motion over the sphere $\mathbb{CP}(1)$ (by
using the complex projective coordinate $\xi=\psi^{1}\diagup\psi^{0}$) as a
single \ complex-valued first order differential equation over the space of
rays:%
\begin{equation}
\dot{\xi}=i\left[  \xi\left(  H_{0}^{0}-H_{1}^{1}\right)  +\xi^{2}H_{1}%
^{0}-H_{0}^{1}\right]
\end{equation}

We find it useful to define the following \textit{real} functions of time in
terms of the matrix elements of the hamiltonian:%

\begin{equation}
\Omega(t)=H_{0}^{0}-H_{1}^{1}\qquad\text{and\qquad}H_{1}^{0}=\overline{H}%
_{0}^{1}=R(t)e^{-i\lambda(t)}
\label{relabeling of the 2-level hamiltonian matrix elements}%
\end{equation}

Note that, though a generic $2\times2$ hermitian matrix has \textit{four}
independent matrix elements, the projection of the motion on $\mathbb{CP}(1)$
needs at most three arbitrary time functions. This is because the projection
of the motion on ray space is invariant under a translation-by-identity
transformation as $\hat{H}\rightarrow\hat{H}+f(t)\hat{I}$ since this would
modify only the overall phase of the state vector but clearly leaves unchanged
the value of $\Omega$. With these three defined parameters together with the
stereographic coordinates provided by (\ref{stereographic map}) we arrive at
the following pair of coupled non-autonomous ODE's over the Bloch sphere:%
\begin{equation}
\left\{
\begin{array}
[c]{c}%
\dot{\theta}=-2R(t)\sin\left(  \varphi-\lambda(t)\right) \\
\dot{\varphi}=\Omega(t)-2R(t)\cot\theta\cos\left(  \varphi-\lambda(t)\right)
\end{array}
\right.  \label{coupled ODE's on CP(1) for theta and phi}%
\end{equation}

For an autonomous system, taking $\left\{  \left\vert u_{0}\right\rangle
,\left\vert u_{1}\right\rangle \right\}  $ as the energy eigenstates $\hat
{H}\left\vert u_{\sigma}\right\rangle =E_{\sigma}\left\vert u_{\sigma
}\right\rangle $ of the hamiltonian, the equation of motion simplifies
drastically to%

\begin{equation}
\left\{
\begin{array}
[c]{l}%
\dot{\theta}=0\\
\dot{\varphi}=\Omega=\text{constant}%
\end{array}
\right.  \qquad\text{with trivial solution\qquad}\left\{
\begin{array}
[c]{c}%
\theta=\theta_{0}\\
\varphi=\Omega t+\varphi_{0}%
\end{array}
\right.
\end{equation}
where $\left\vert u_{0}\right\rangle $ and $\left\vert u_{1}\right\rangle $
represent respectively the north and south poles of the Bloch sphere. The
possible time-evolutions are the uniform circular motions around the pole-axis
with constant angular velocity $\Omega=E_{0}-E_{1}$. For an arbitrary
\textit{non-autonomous} system we may diagonalize \textit{instantaneously} the
hamiltonian $\hat{H}(t)$, finding \textit{time-dependent} energy eigenkets
$\left\{  \left\vert E_{0}(t)\right\rangle ,\left\vert E_{1}(t)\right\rangle
\right\}  $ represented by the antipodal points $P_{0}(t)$ and $P_{1}(t)$ on
the Bloch sphere $S^{2}$.
\begin{figure}[H]
\begin{center}
\includegraphics[scale=.4]{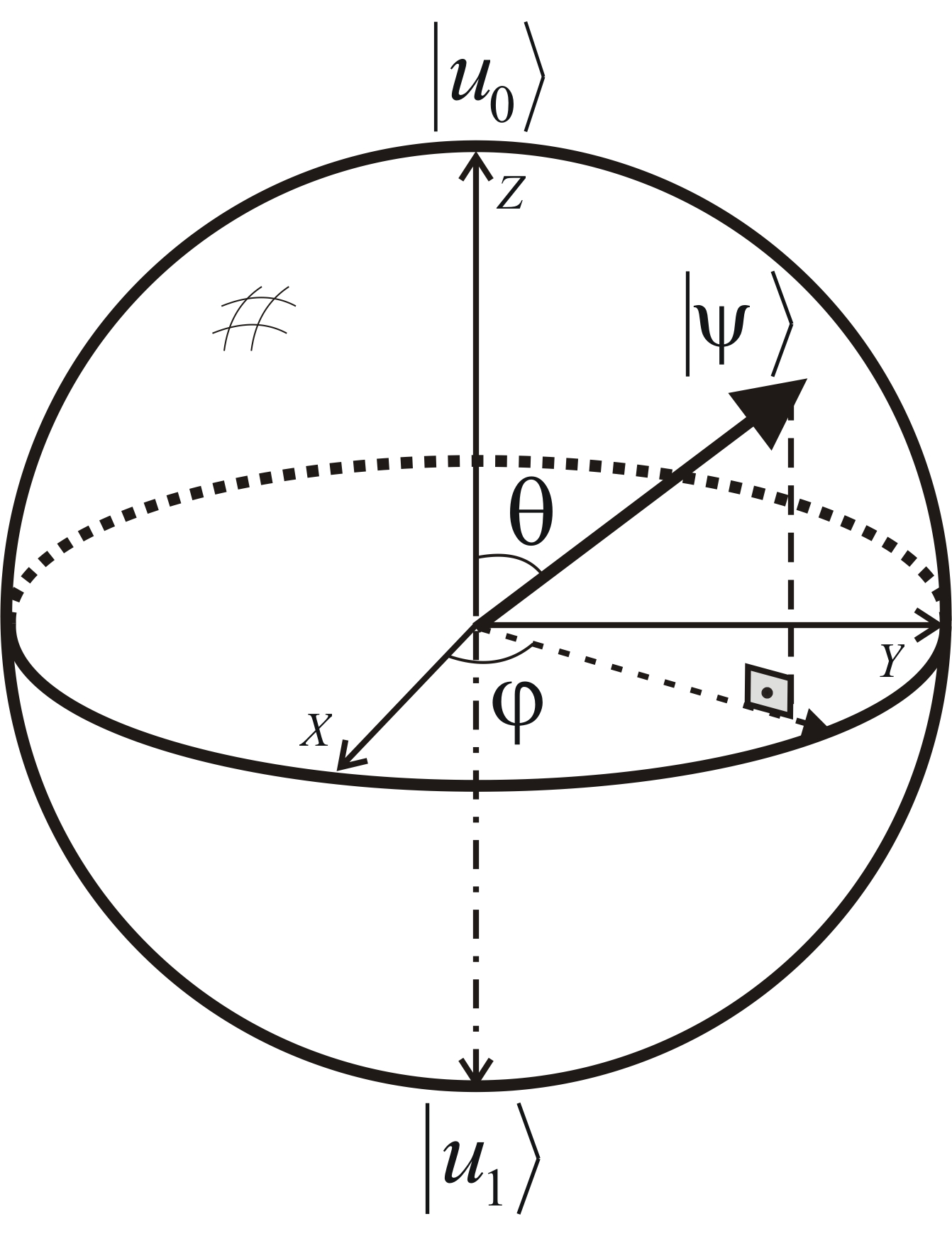}
\end{center}
\caption{Bloch sphere}%
\end{figure}

The motion is now given instantaneously by a circular motion in turn of an
instantaneous axis determined by $P_{0}(t)$ and $P_{1}(t)$ with instantaneous
angular velocity
\begin{equation}
\omega(t)=E_{0}(t)-E_{1}(t), \label{instantaneous angular velocity}%
\end{equation}

We may introduce coordinates $\theta\prime(t)$ and $\varphi\prime(t)$ for the
point $P_{0}(t)$ by the stereographic map%
\begin{equation}
\tan\left(  \theta\prime/2\right)  e^{i\varphi\prime}=(\left\langle
u^{0}\right\vert E_{0}(t)\rangle)^{-1}\left\langle u^{1}\right\vert
E_{0}(t)\rangle.
\end{equation}
The right side of the above equation can be computed by the eigenvalue
equation:%
\begin{equation}%
{\displaystyle\sum\limits_{\sigma}}
\left\langle E^{\tau}\right\vert \mu_{\sigma}\rangle(H_{\nu}^{\sigma}-E^{\tau
}\delta_{\nu}^{\sigma})=0, \label{eigenvalue equation}%
\end{equation}
giving us%
\begin{equation}
\left\{
\begin{array}
[c]{c}%
E^{0}=\frac{1}{2}\left[  H_{0}^{0}+H_{1}^{1}+\left(  \Omega^{2}+4R^{2}\right)
^{1/2}\right] \\
E^{1}=\frac{1}{2}\left[  H_{0}^{0}+H_{1}^{1}-\left(  \Omega^{2}+4R^{2}\right)
^{1/2}\right]
\end{array}
\right.
\end{equation}
We make now the following identifications%
\begin{equation}
\left\{
\begin{array}
[c]{l}%
\lambda(t)=\varphi\prime(t)\\
R(t)=\dfrac{1}{2}\omega(t)\sin\theta\prime(t)\\
\Omega(t)=\omega(t)\cos\theta\prime(t)
\end{array}
\right.
\end{equation}
and we arrive at a very convenient form for our coupled ODE's over the sphere
$\mathbb{CP}(1)$:%
\begin{equation}
\left\{
\begin{array}
[c]{l}%
\dot{\theta}=\omega(t)\sin\theta\prime(t)\sin(\varphi-\varphi\prime(t))\\
\dot{\varphi}=\omega(t)\left[  -\cos\theta\prime(t)+\sin\theta\prime
(t)\cot\theta\cos(\varphi-\varphi\prime(t))\right]  .
\end{array}
\right.  \label{Coupled EDO's over the sphere 2}%
\end{equation}

The three time dependent parameters that describe our hamiltonian can now be
directly related to the geometry of the Riemann sphere: $\omega(t)$ gives
us the instantaneous angular velocity of the motion in turn of the
instantaneous axis of revolution whose direction is described by
$(\theta\prime(t),\varphi\prime(t))$. Surprisingly, this seemingly abstract
formulation is actually implemented by nature as the motion of an electron
spin immersed in a time dependent classical magnetic field given by (see
\cite{sakurai1994}, for instance)
\[
\vec{B}(t)=\frac{m}{e}\omega(t)\left[  \sin\theta\prime(t)\left(  \cos
\varphi\prime(t)\hat{\imath}+\sin\varphi\prime(t)\hat{\jmath}\right)
+\cos\theta\prime(t)\hat{k}\right]  \qquad\text{(with }\hslash=c=1\text{)}%
\]
Let us consider now a very specific kind of motion where the hamiltonian (the
point $P_{0}(t)$) describes a uniform rotation on a great circle of the sphere
which we choose to be the equator without any loss of generality:%
\[
\theta\prime(t)=\pi/2\qquad\varphi\prime(t)=\Omega t\qquad\text{and\qquad
}\omega=\omega_{0}=\text{constant}%
\]
Equation (\ref{Coupled EDO's over the sphere 2}) becomes then%
\begin{equation}
\left\{
\begin{array}
[c]{l}%
\dot{\theta}=\omega_{0}\sin(\varphi-\Omega t)\\
\dot{\varphi}=\omega_{0}\cot\theta\cos(\varphi-\Omega t)
\end{array}
\right.  \label{coupled EDO's on the sphere 3}%
\end{equation}
We choose initial conditions%
\begin{equation}
\theta(0)=0\qquad\text{and\qquad}\varphi(0)=0 \label{initial conditions}%
\end{equation}
to match the initial state vector as an energy eigenket. The above initial
conditions together with (Eq. \ref{coupled EDO's on the sphere 3}) lead to%
\begin{equation}
\dot{\theta}(0)=0\qquad\text{and\qquad}\dot{\varphi}(0)=0
\label{initial conditions 2}%
\end{equation}

To prove the adiabatic theorem for this special case, all we need is to assure
that for a \textit{sufficiently small value} of $\Omega/\omega_{0}$ the
conditions
\[
\theta(t)\approx\pi/2\qquad\text{and\qquad}\varphi(t)\approx\Omega t
\]
must hold in a \textit{self-consistent} manner. Indeed, given that
$\varphi\approx\Omega t$, equation (\ref{coupled EDO's on the sphere 3})
implies that $\theta\approx\operatorname{arccot}\left(  \Omega/\omega
_{0}\right)  \approx\operatorname{arccot}\left(  0\right)  =\pi/2$. To prove
the contrary implication, assume $\theta\approx\pi/2$. It is convenient then
to introduce new coordinates
\begin{equation}
\left\{
\begin{array}
[c]{l}%
\varepsilon(t)=\theta(t)-\pi/2\\
\eta(t)=2(\varphi-\Omega t)
\end{array}
\right.  \label{new coordinates}%
\end{equation}
which gives us
\begin{equation}
\left\{
\begin{array}
[c]{l}%
\dot{\varepsilon}=\omega_{0}\sin\left(  \eta/2\right) \\
\dot{\eta}=-2\left[  \Omega+\omega_{0}\tan\varepsilon\cos\left(
\eta/2\right)  \right]
\end{array}
\right.  \qquad\text{with\qquad}\left\{
\begin{array}
[c]{l}%
\varepsilon(0)=0\\
\eta(0)=0
\end{array}
\right.  \qquad\text{and}\qquad\left\{
\begin{array}
[c]{l}%
\dot{\varepsilon}(0)=0\\
\dot{\eta}(0)=-2\Omega
\end{array}
\right.  \label{coupled EDO's 4}%
\end{equation}
By taking the time derivative of the second equation (\ref{coupled EDO's 4})
and using the first equation together with the condition $\theta(t)\approx
\pi/2\Longrightarrow\varepsilon\approx0$ one gets the approximate differential
equation
\begin{equation}
\ddot{\eta}\approx-\omega_{0}^{2}\sin\eta\label{pendulum-like equation}%
\end{equation}
which we recognize immediately as the well-known simple pendulum equation
(from undergraduate classical mechanics) with ``conserved energy" given by
\begin{equation}
E=T+U=\frac{1}{2}\dot{\eta}^{2}-\omega_{0}^{2}\cos\eta
\label{energy for classical pendulum}%
\end{equation}
It follows from the initial conditions in (\ref{coupled EDO's 4}) that the
\textquotedblleft total mechanical energy" of the pendulum is
\begin{equation}
E=2\Omega^{2}-\omega_{0}^{2} \label{total energy of pendulum}%
\end{equation}
with minimum\ \textquotedblleft potential energy" at $U(\eta_{\min
})=U(0)=-\omega_{0}^{2}$. It is easy to see in this set up that one can
confine the motion as close as one desires to $\eta_{\min}=0$ by making the
ratio $\omega_{0}/\Omega$ as \textit{larger} as necessary. So we
have that the implication $\theta\approx\pi/2\Longrightarrow\varphi
\approx\Omega t$ must hold true. 
\begin{figure}[ptb]
\begin{center}
\includegraphics[scale=.2]{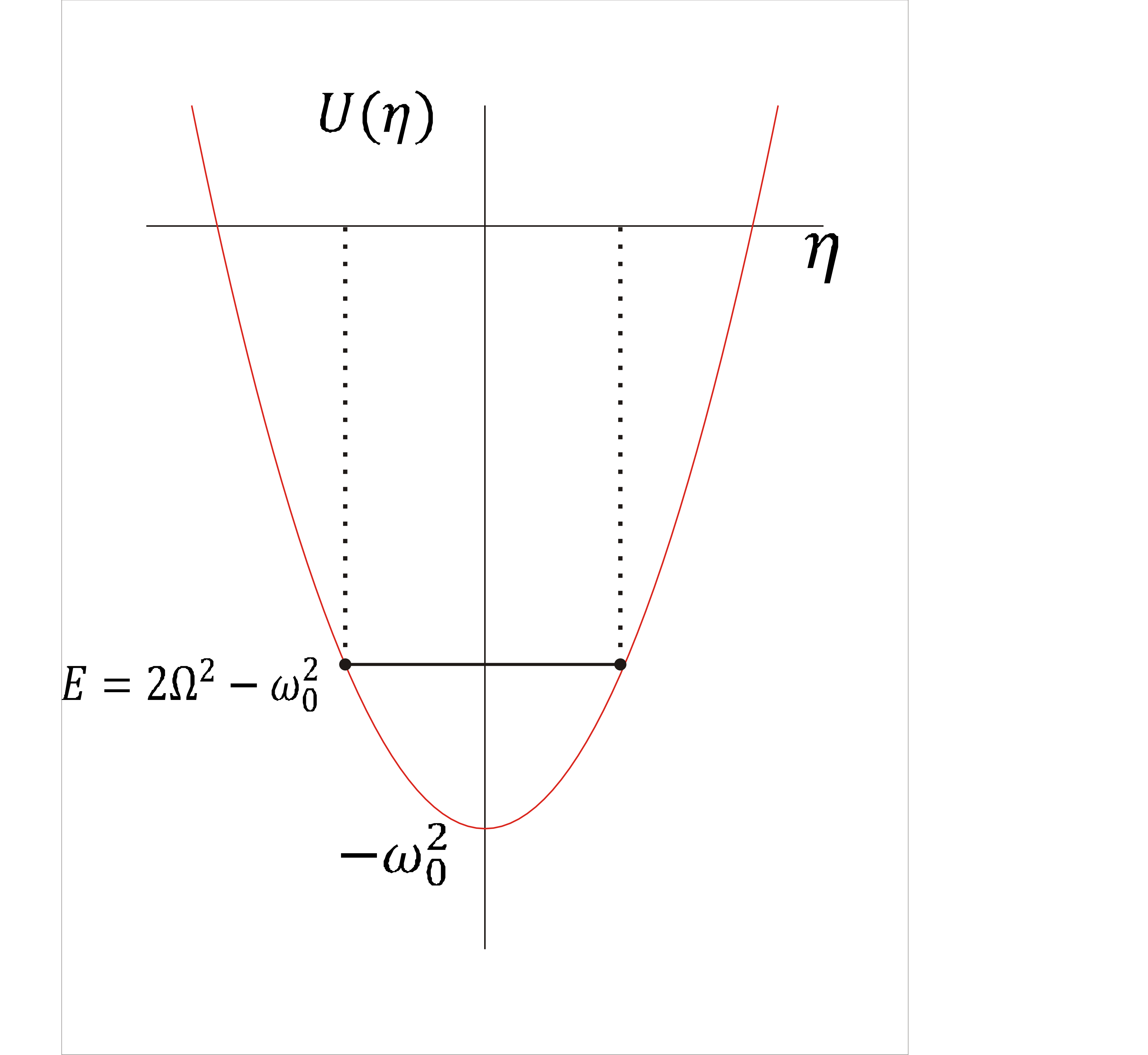}
\end{center}
\caption{Potential energy.}%
\end{figure}

To extend the proof to an arbitrary path over the sphere one must only need to
remember that such path can be divided in $N$ (let us say equal segments)
sub-paths with $N$ sufficiently large so that each segment may be
considered as approximately a geodesic path in such a way that the above
argument can be applied piece-wise. 
\begin{figure}[ptb]
\begin{center}
\includegraphics[scale=.4]{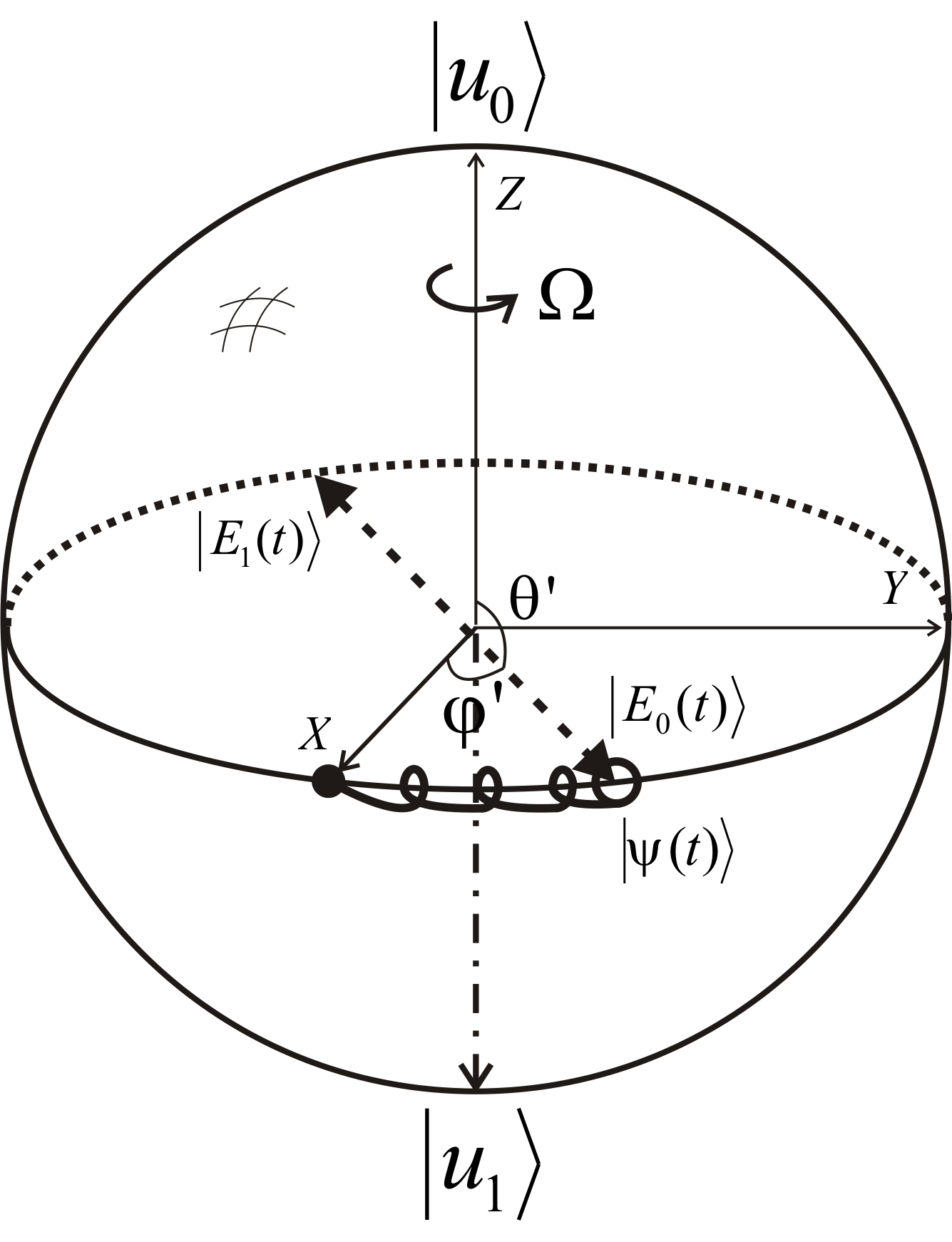}
\end{center}
\caption{A pictorial representation of our simplified model of the adiabatic
theorem in action.}%
\end{figure}

We have plotted below some numerical implementations of our simplified model (Eq. \ref{coupled EDO's on the sphere 3}) with different values of $\omega_{0}/\Omega$:
\begin{figure}[H]
\begin{center}
\includegraphics[scale=0.4]
{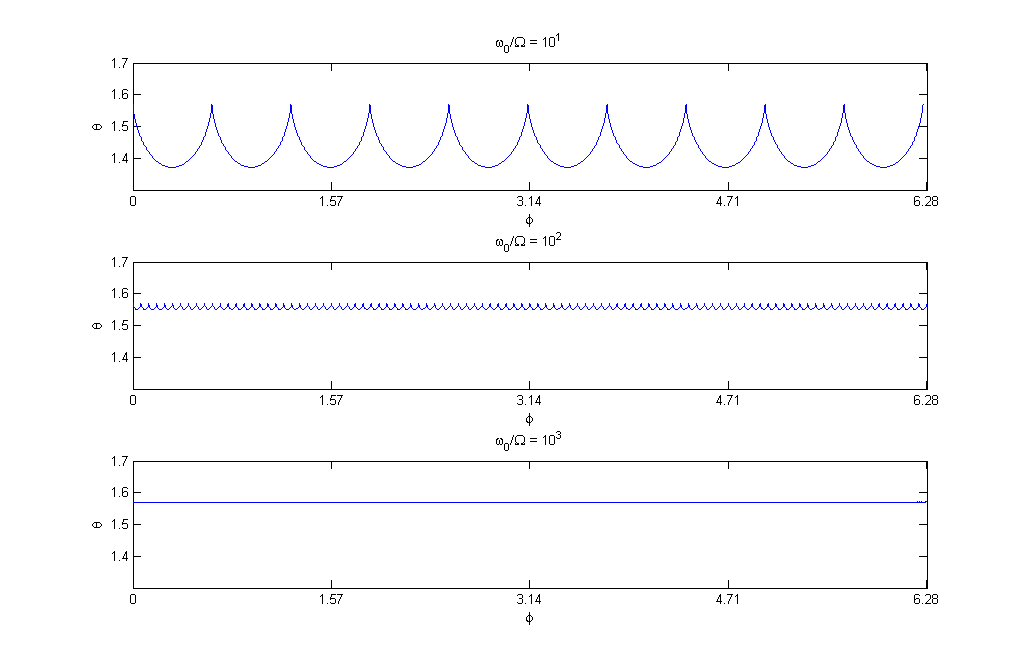}
\end{center}
\caption{Three plots for $\omega_{0}/\Omega$ respectively given by
$10^{1},10^{2}$ and $10^{3}$.}%
\end{figure}The above analysis shows convincingly that the projected state
``whirls" in turn of the instantaneous north pole evermore tightly as the
factor $\omega_{0}/\Omega$ gets larger and larger.

\section{Concluding remarks}

We have briefly reviewed the geometry of quantum mechanics developed back in
the eighties by Aharonov \ and Anandan among others. We used these ideas to
present an elementary geometric argumentation for the validity of the
adiabatic theorem. We have used only the minimum geometric structure needed
for an understanding of the adiabatic theorem, but there are many rich
structures in the geometry of quantum mechanics. We refer the reader to
\cite{arnold1989mathematical,
kibble1979geometrization,page1987,brody2001geometric} for more details on
these matters.

Though we do not claim this as a complete proof, we are sure that this
geometric reasoning can be easily implemented as a rigorous proof for those
analytically inclined. We hope that this geometric picture is sufficiently
straightforward to turn the concept of quantum adiabatic computation into a
much more understandable tool for undergraduate students of physics and even
for specialists.

The modern theories of quantum computation and information have become a
central research field in physics and the quest for the best models of quantum
computing (quantum gate model, adiabatic quantum computing, cluster or
measurement-based quantum computation) and even to establish the equivalence
(or not) between their true powers of computation (computability and
complexity classes) are still open questions. A better intuitive understanding
and presentation of the key physical and mathematical structures behind their
functioning seems to be a sensible enterprise. We believe to have presented
the most direct and pedagogical frame so far of this important theorem, at
least for two level systems.

\section*{Acknowledgments}

A. C. Lobo wishes to acknowledge financial support from
\textit{NUPEC-Funda\c{c}\~{a}o Gorceix}, C. A. Ribeiro, P. R. Dieguez and R.
A. Ribeiro acknowledge financial support from \textit{Conselho Nacional de
Desenvolvimento Cient\'{\i}fico e Tecnol\'{o}gico} (CNPq). The authors also
wishes to thank the anonymous referee for many valuable suggestions that
helped much to improve the quality of this manuscript. The authors thank the
assistance with the English language from Fernanda Lobo Bellehumeur and we
also thank D. Brody for the references on the geometry of quantum mechanics.

{\normalsize
\bibliographystyle{plain}
\bibliography{referencias}
}

\end{document}